# Signaling Network Assessment of Mutations and Copy Number Variations Predicts Breast Cancer Subtype-specific Drug Targets

Naif Zaman,[1,2,3,6] Lei Li,[4,6] Maria Luz Jaramillo,[1] Zhanpeng Sun,[5] Chabane Tibiche,[1] Myriam Banville,[1] Catherine Collins,[1] Mark Trifiro,[5] Miltiadis Paliouras,[5] Andre Nantel,[1,2] Maureen O'Connor-McCourt,[1] and Edwin Wang [1,3,*]

[1]National Research Council Canada, Montreal, Canada

[2]Department of Anatomy and Cell Biology, McGill University, Montreal, Canada

[3]Center for Bioinformatics, McGill University, Montreal, Canada

[4]Department of Molecular Genetics, University of Toronto, Toronto, Canada

[5]Department of Experimental Medicine, McGill University, Montreal, Canada

[6]These authors contributed equally to this work
[*]Correspondence: edwin.wang@cnrc-nrc.gc.ca

# SUMMARY

Individual cancer cells carry a bewildering number of distinct genomic alterations (i.e., copy number variations and mutations), making it a challenge to uncover genomic-driven mechanisms governing tumorigenesis. Here we performed exome-sequencing on several breast cancer cell lines which represent two subtypes, luminal and basal. We integrated this sequencing data, and functional RNAi screening data (i.e., for identifying genes which are essential for cell proliferation and survival), onto a human signaling network. Two subtype-specific networks were identified, which potentially represent core-signaling mechanisms underlying tumorigenesis. Within both networks, we found that genes were differentially affected in different cell lines; i.e., in some cell lines a gene was identified through RNAi screening whereas in others it was genomically altered. Interestingly, we found that highly connected network genes could be used to correctly classify breast tumors into subtypes based on genomic alterations. Further, the networks effectively predicted subtype-specific drug targets, which were experimentally validated.

# INTRODUCTION

Thus far, several thousands of tumors representing more than 20 cancer types have been sequenced. These efforts have identified thousands of genomic alterations such as somatic mutations, amplifications, deletions, chromosomal translocations and gene-fusions in each individual cancer genome (Banerji et al., 2012; Stephens et al., 2012; Cancer Genome Atlas Network, 2012). With so many genomic alterations in each tumor genome, it is a big challenge to dissect, prioritize and uncover the functional importance of the genomic alterations and the underlying mechanism that drive cancer development, progression and metastasis (Chin et al., 2011).

During cancer cell evolution, some genomic alterations become the underlying cause for tumor cell proliferation, fitness and clonal selection. Cell survival, proliferation and apoptosis are the most primitive and fundamental cancer hallmarks (Hanahan and Weinberg, 2011). Systematic identification of genes which are essential for cell proliferation and survival or cancer essential genes (i.e. functional screens in which gene knocking-down results in cancer cell growth inhibition) by genome-wide RNAi screening has showed that indeed there exist distinct subsets of genes that are selectively required by different cancer cells (Schlabach et al., 2008; Silva et al., 2008). These results suggest that different cancer cells have unique growth and survival requirements which are acquired through distinct subsets of genomic alterations. For a more comprehensive understanding of how genomic alterations drive cancer cell survival and proliferation, it is imperative to conduct an analysis that integrates genomic alteration information and functional genetic data (i.e., via genome-wide RNAi screenings) from the same individual cancer cell. Tumors are highly heterogeneous, sequencing of different regions of a tumor generated different sets of mutations (Gerlinger et al., 2012). The mixture of mutations in a tumor prevents linking genes to functions. It could be interesting to dissect and sequence the major clones of tumors or conduct single-cell genome sequencing so that each mutation could be functionally investigated in the cell/clone which bears that mutation. Toward this end, in this study we performed genome-wide exome-sequencing for a panel of breast cancer cell lines and matched their corresponding genome-wide

RNAi screening data (Marcotte et al., 2012) to perform an integrated network analysis to get insights into the underlying mechanisms of cancer cell survival and proliferation driven by genomic alterations.

Breast cancers have been classified into three molecular subtypes: luminal, HER2 and basal (basal A and basal B) (van 't Veer et al., 2002), using a 50-gene expression signature (PAM50) (Parker et al., 2009). The HER2 subtype often has ERBB2 mutated or amplified and has had some degree of clinical success because of effective therapeutics that can target ERBB2 (Slamon et al., 1987). The luminal subtype is often characterized by the expression of estrogen receptor (ER+), which is not expressed in the basal subtype. The ER+ group (known as luminal breast cancer) has some degree of varying drug response, while triple-negative breast cancers (known as basal-like breast cancer) lack the expression of ER, progesterone receptor and HER2 and have very limited chemotherapy or other molecularly targeted drugs available. Therefore, we focused on developing integrated networks, composed of both genetic screening (RNAi screening) and genomic alteration (mutation and copy number variations) data, to further characterize the luminal and basal breast cancer subtypes. This novel approach is likely to generate more insights into the fundamental network wiring in cancer with the more focused aim to identify subtype-specific breast cancer genes which may lead to better treatment options in the near future.

**RESULTS AND DISCUSSION**

**Genome Sequencing of Breast Cancer Cell Lines**

A genome-wide cell survival RNAi screen has previously been conducted for a panel of luminal and basal breast cancer cell lines (Marcotte et al., 2012). Furthermore, since five lines in the panel have already been exome-sequenced (Sjoblom et al., 2006), we performed an exome-sequencing of the remaining 11 lines (see Extended Experimental Procedures; Table S1). After removing naturally occurring genetic polymorphisms using the data from dbSNP database and 1000 genome project (see Extended Experimental

Procedures), we identified 3,817 somatic point mutations. Of these tumor associated genetic alterations, 2,548 were predicted to generate missense mutations (annotated as non-synonymous mutation by Annovar (http://www.openbioinformatics.org/annovar/)), 192 nonsense (or stopgain) mutations, 111 mutations were shown to contain an essential splice site, 4 mutations resulted in stop codon read-through (or stopless) mutations, and 1,073 were synonymous substitutions which would result in silent changes in protein sequence. We also identified, 164 small insertions or deletions (79 and 85, respectively), of which 94 introduced translational frameshifts while 50 were in-frame, 5 stopgain and 1 stoploss SNVs (Table S2). Based on the Annovar program which predicts potential functional mutations, we obtained 1,630 potential driver-mutating genes (i.e., cancer-causing genes) for all the 11 cell lines (Table S3). Mutants of MAP kinase family were found across all of the lines. As expected, mutant TP53 (80%) was associated with basal subtypes. These results are in agreement of the results of genome sequencing of nearly 1,000 breast cancer samples (Banerji et al., 2012; Stephens et al., 2012; Cancer Genome Atlas Network, 2012). We also compared the driver-mutating genes in this study to those derived from COSMIC database and ~1,000 breast cancer samples mentioned above, and found 45 novel driver-mutating genes in at least one cell line. Three genes ZBTB18, TENM4, TMEM178A (Table S3) among them are found in two cell lines.

**Subtype-Specific Survival Signaling Networks Highlight the Evolutionary Convergence of Selective Genomic Alterations**

Cells employ signaling pathways and networks to drive biological processes. Genomic alterations in signaling pathways and networks might result in malignant signaling which then leads to cancer phenotypes. Genome-wide RNAi screening experiments not only uncover cancer essential genes, but also pinpoint genes which are involved in influencing cell proliferation. Knocking-down a proliferation influencing gene will not necessarily lead to cell death, but will greatly reduce cancer cell growth (see Experimental Procedures, Fig S1). If a gene which is involved in the regulation of proliferation genes is also subject to nonsynonymous genetic alterations (mutations) or amplified, we defined that gene as a 'cell survival related driving-regulator' (called 'driving-regulator' in this study) (see Experimental Procedures, Fig S1). Previously we showed that modeling and perturbing of signaling networks (Rozenblatt-Rosen et al., 2012; Cui et al., 2007; Barabasi et al., 2011) and cancer hallmarks (Li et al., 2010) led to getting insights into cancer gene mutations and identifying high-quality cancer biomarkers. To obtain further insight into the underlying mechanism of cancer cell proliferation trigged by cancer genomic alterations, for each cell line we mapped driving-regulators, cancer essential and proliferation influencing genes onto a manually curated human signaling network (containing ~6,000 genes and 50,000 relations) (Awan et al., 2007; Cui et al., 2007; Li et al., 2012) to generate integrated cell line-specific survival networks (Figure 1; see Experimental Procedures). Such a network represents the signaling mechanism for cancer cell survival and proliferation. The gene amplification data processed using GISTIC (Beroukhim et al., 2007) were obtained from Cancer Cell Line Encyclopedia (CCLE, http://www.broadinstitute.org/ccle/home). Detailed information for defining cancer essential and proliferation influencing genes, and 'driving-regulators' are explained in Figure S1 and Experimental Procedures.

Highly connected network genes, called hubs, act as global signal integrators or global regulators for multiple signaling pathways (Wang et al., 2007; Wang, 2010). To find out whether the driving-regulators, and essential and proliferation influencing genes shape the survival networks, we conducted both fuzzy k-mean clustering and hierarchical clustering analyses of the cell lines using the hubs of cancer essential genes, driving-regulators or both, respectively. In this study, we defined the top 10% of highly connected genes in a network as hubs. In general, we also tested the hubs using top 15% as a cutoff in all the analyses and found both cutoffs generated similar results. As seen in Figure 2A and 2B, the hubs of either driving-regulators ($p = 0.12$, fuzzy k-mean clustering and Fisher's test) or essential genes ($p = 1.0$, fuzzy k-mean clustering and Fisher's test) alone were unable to classify the individual cell lines to the luminal and basal subtypes. However, when we combined the hubs of the driving-regulators and essential genes, the cell lines were better classified and distinguished into the luminal and basal subtypes (Figure 2C, $p = 0.03$, fuzzy k-mean clustering and Fisher's test). Permutation tests (see Extended Experimental Procedures) showed that significant classification of luminal and basal subtypes by the network hubs couldn't be obtained by random ($p = 9.0 \times 10^{-4}$). These results suggest that although both driving-regulators and essential genes are profoundly different between cancer cells (Figure S2), they are complementary and converge to similar survival signaling mechanisms within their respective subtype. To further explore these observations in detail, we constructed subtype-specific survival networks (see Extended Experimental Procedures). A subtype-specific network contains ~200 genes which appear across ≥50% of a subtype's cell lines. Nearly all the genes (>95%) in a subtype-specific network, act as cancer essential genes in one cell line but as driving-regulators in another line (Figure 3). Randomization tests (see Extended Experimental Procedures) showed that the recurrent usage of the genes in luminal and basal subtypes, respectively, is not by random ($p < 1.0 \times 10^{-4}$). These network genes are recurrently used by the subtype's lines, suggesting that cancer cells are 'addicted' to their respective subtype-specific network for survival and proliferation. A subtype-specific network represents core survival signaling mechanisms which shed light on convergent evolutionary events, and provides functional constraints for selecting genomic alterations that could offer a competitive growth advantage for cancer cells. The selective pressure led to the emergence of distinct network hubs in the luminal and basal subtypes (Figure 3A, 3B and 3C). For example, AKT1, PIK3 and ESR1 are dominantly selected in luminal, whereas TP53 and SRC are genetically dominant in the basal subtypes.

We explored network modules for the subtype-specific networks using the Gene Ontology-guided Markov Cluster (MCL) Algorithm (see Extended Experimental Procedures). Three functional modules where one is centered by CDK1 for cell cycle, one is centered by P53 for apoptosis and genome instability and another one is centered by growth factors such as EGFR and MAPK pathway components for cell proliferation, were found in the basal-specific networks (Figure 3A and 3B). Two network modules, where one is centered by CDK1/MYC for cell cycle, and the other is centered by AKT/PIK3CA, growth factors such as MET and MAPK pathway components for cell proliferation and growth, were found in the luminal-specific network (Figure 3C). To further interpret these findings, we conducted pathway enrichment analysis (see Extended

Experimental Procedures) for each subtype-specific network using the cancer essential genes, proliferation influencing genes and driving-regulators. Signaling pathways of cell cycle, apoptosis, MAPK/growth factors (i.e., MET) and transcription processes were found in both luminal and basal lines, highlighting the fact that these cancer subtypes share core survival pathways commonly used by breast cancers (Table S4). In addition, cancer cells of the basal subtype (basal A and basal B) share the signaling pathways for genome instability such as P53, DNA repair, and telomere extension and maintenance (Table S4), which were not commonly used by luminal cells. Most of the essential genes affecting genome instability pathways are relatively unique for the basal subtype, which highlights the signature of basal subtype, and provides unique drug targets for the aggressive groups such as triple negative groups.

**Subtype-Specific Survival Signaling Networks Provide Predictive Power**

The convergence of the cancer essential genes and driving-regulators into their respective subtype-specific survival networks suggests that in each subtype there is a 'deterministic' path for cancer cell proliferation driven by genomic alterations, and therefore, the networks could provide 'predictable' power for selective genomic alterations. Consequently, we tested whether the integrated subtype-specific networks could have predictive power in order to accurately identify breast cancer tumor subtypes. To demonstrate this, we used the hub genes of the subtype-specific networks to classify the 16 cell lines. To do so, we first identified differential hubs between the subtype-specific networks (see Extended Experimental Procedures) and then classify the 16 cell lines. Indeed, hub genes were able to distinguish between luminal and basal subtypes (Figure 4A, $p = 1.2 \times 10^{-4}$, fuzzy k-mean clustering and Fisher's test). These results suggest that amplification or mutation of a few top hub genes could activate the entire network for cancer cell survival and proliferation. Therefore, we extended this analysis to demonstrate that these hub genes' genomic alteration profiles (amplification and drive-mutating status) were able to significantly classify 402 breast tumor samples (see Extended Experimental Procedures) into the basal and luminal subtypes (Figure 4B, $p = 2.2 \times 10^{-16}$, fuzzy k-mean clustering and Fisher's test). These results highlight the convergent and deterministic properties of selective genomic alterations, which exploit distinct core survival signaling networks (i.e., subtype-specific networks) for cancer cell proliferation. These genomic alterations could be gradually or suddenly (i.e., through genome duplication) accumulated and then be selected during cancer evolution. Detection of the genomic alterations of a fraction or all of the genes in this hub gene set could help in early diagnosis of breast tumors. Recently, plasma genome sequencing approach has been showed that copy number variations and mutations of plasma DNA are detectable and comparable between cancer and healthy individuals (Chan et al., 2012; Leary et al., 2012). As sequencing costs continue to decline, these genes could be used to develop non-invasive tests (e.g., using plasma genome sequencing (Murtaza et al., 2013)) for screening very early-stage breast cancer patients or distinguish breast cancer subtypes.

To further demonstrate their predictive power of the subtype-specific networks, we seek out to predict subtype-specifically therapeutic interventions. If a hub gene specifically appears in either luminal or basal subtype-specific networks, we expected that that gene could be a drug target specifically for its subtype. Based on this criterion, AKT1, mTOR, MET, MDM2, HSP90AA1, RAF1, SFN, FYN, CHEK1, ESR1 were predicted as potentially luminal-specific drug targets, while TGF-beta, IGF1R, MAPK3, GRB2, SRC, TUBB, JAK2 and EGFR were predicted as potentially basal-specific drug targets (undruggable differential hubs between subtypes such as transcription factors like P53 were not considered). To validate these predictions, we obtained the data from systematic drug screenings of cancer cell lines including over 40 breast cancer cell lines (Garnett et al., 2012; Heiser et al., 2012) and statistically evaluated the sensitivity of these drugs for luminal and basal subtypes (see Experimental Procedures). The predicted targets, among which have been included in the drug screenings, except MAP2K1 and CHEK1 were in agreement with the experimentally screening results (Table 1).

In summary, using an integrative network analysis of the data derived from exome-sequencing and genome-wide RNAi screening of a breast cancer cell line panel, we have shown that a set of primitive core signaling pathways such as cell cycle, apoptosis, growth factors/MAPK and transcription is commonly exploited by genomic alterations for cancer cell survival in all the breast cancer cells, while the signaling pathways of P53 and genome instability such as telomere maintenance are specifically exploited by genomic alterations in the basal subtype. The essential genes in these pathways are unique drug targets for the aggressive breast (i.e., basal subtype) cancer groups. The functional convergence of the essential genes and driving-regulators in a limited number of signaling pathways lead to emerging of subtype-specific survival signaling networks in which genes recurrently switch roles between cancer essential genes and cancer driving-regulators in cancer cells. These networks elucidate underlying signaling mechanisms governing cancer cell survival and proliferation, and imply selective pressures for evolutionary convergence of cancer genomic alterations. However, it is clear that signaling mechanisms of the two subtypes are different, which is evident by the fact that a set of network genes (i.e. genes are differentially different between the two subtype-specific networks) whose genomic alteration profiles (amplification and mutating status) are able to significantly distinguish breast tumor samples into luminal and basal subtypes, and furthermore, these networks predicted subtype-specific drug targets. Importantly, most (~80%) of the predicted drug targets have been experimentally validated. Together with the finding that more amplified genes could act as cancer drivers, these results may have profound clinical implications in personalized treatment of cancer patients (Wang et al., 2013a; Wang et al., 2013b), and screening early breast cancer patients by plasma DNA sequencing using the set of network genes.

**EXPERIMENTAL PROCEDURES**

**Samples for Exome-Sequencing**

Eleven breast cancer cell lines (i.e. BT549, MDAMB436, BT20, MDAMB231, MDAMB468, SKBR3, ZR751, HCC1500, MDAMB453, MCF7 and T47D) were obtained from ATCC for exome-sequencing.

**Datasets**

Exome-sequencing data for five breast cancer cell lines (Table S1) were obtained from Sjoblom et al. (Sjoblom et al., 2006). Microarray and copy number data of the sixteen breast cancer cell lines were obtained from the Cancer Cell Line Encyclopedia (CCLE, http://www.broadinstitute.org/ccle/home). Data for genome-wide RNAi screening of cell survival and proliferation of the 16 breast cancer cell lines were obtained from COLT-Cancer database (http://colt.ccbr.utoronto.ca/cancer/). The human signaling network (Version 4, containing ~6,000 genes and ~50,000 relations) includes data from the manually curated signalling networks by us previously (Awan et al., 2007; Cui et al., 2007; Li et al., 2012) and by PID (http://pid.nci.nih.gov/) and our recent manual curations using iHOP database (http://www.ihop-net.org/UniPub/iHOP/). Pathway gene lists were obtained from GSEA Molecular Signatures Database (http://www.broadinstitute.org/gsea/msigdb/index.jsp). Data of systematic drug screenings of breast cancer cell lines were obtained from these studies (Garnett et al., 2012; Heiser et al., 2012).

**Cancer Essential Genes, Proliferation Influencing Gene and Driving-Regulators**

The descriptions (below) of driving-regulators, essential and proliferation influencing genes are summarized in Figure S1. Genome-wide RANi screening results of the 16 breast cancer cell lines was collected in COLT-Cancer database (http://colt.ccbr.utoronto.ca/cancer/). In the database, the essentiality of each gene for a given cell line has been scored based on GARP (Gene Activity Rank Profile) scores and P values, which were computed in each experiment of the genome-wide RANi screening (Marcotte et al., 2012). Lower P value depicts higher significance for the 'higher gene essentiality' (e.g., higher degrees of influencing cell survival). Details for calculating of GARP scores and P value were described previously (Marcotte et al., 2012). Housekeeping genes were also annotated in the database. If a gene in a given cell line has a RANi-screening p value less than 0.05 and is not belonging to housekeeping genes, that gene was defined as a 'cancer essential gene' in that cell line (Marcotte et al., 2012). Validation experiments supported this p value cutoff (i.e., 0.05) for defining the cancer essential genes (Marcotte et al., 2012). If a gene in a given cell line has a RANi-screening p value less than 0.1 but greater than 0.05, we defined that gene as a 'proliferation influencing gene' in that cell line. We assumed that knocking-down a proliferation influencing gene will not lead to cell death, but significantly reduce cell growth and survival. We asked that an essential gene, proliferation influencing gene in a given cell line should be among the top 75% of the expressed genes for that cell line as described previously (Cancer Genome Atlas Network, 2011). Amplification genes are considered if they have a GISTIC score > 0.3 and are among the top 50% of the expressed genes for that cell line. The cutoff 0.3 is widely used to define gene amplifications (Mermel et al., 2011; Barretina et al., 2012). Details of setting the cutoff of 50% for gene expression are explained in Extended Experimental Procedures. If an amplified gene in a given cell line has a RANi-screening p value less than 0.4, we defined this gene as a cell survival related 'driving-regulator' in that cell line, assuming that knocking-down the driving-regulators will affect cell growth and survival. It should be noted that the definitions of these terms are based on certain cutoffs. We changed the cutoffs of RANi-screening P values for these genes (i.e., $p < 0.03$, $0.03 < p < 0.1$ and $p < 0.5$ for 'cancer essential genes', 'proliferation influencing genes', and 'driving-regulators', respectively) and re-ran all the analyses in this study. We found that the results are similar to those obtained using the original cutoffs. However, we still like to warn that interpreting of the results should take in consideration of the definitions and the cutoffs used in this study.

**Drug Sensitivity Analysis**

For a given drug, we compared the IC50 values between luminal and basal lines. Kruskal–Wallis ANOVAs were used to evaluate the statistically significant differences of the IC50 values between the subtypes. Heiser et al. (Heiser et al., 2012) performed drug screening on more breast cancer cell lines (~50 cell lines) than Garnett et al. (Garnett et al., 2012). Therefore, we mainly used the data from Heiser et al.

**ACKNOWLEDGEMENTS**

This work is supported by NRC's Genome and Health Initiative, Genome Canada and Canadian Institutes of Health Research. We thank Dr. Z. Yu at Lady Davis Institute for discussing the manuscript.

**Table 1. Validation of the Predicted Subtype-Specific Drug Targets**

| Compound | Predicted subtype-specific drug target | Basal vs Luminal (P value) | Subtype specificity |
|---|---|---|---|
| SIGMA AKT12 INHIBITOR | AKT1, AKT2 (luminal) | 5.04E-04 | Luminal |
| TAMOXIFEN | ESR1 (luminal) | 3.92E-02 | Luminal |
| NUTLIN3A | MDM2 (luminal) | 3.13E-02 | Luminal |
| RAPAMYCIN | mTOR (luminal) | 1.78E-03 | Luminal |
| 17-AAG | HSP90 (luminal) | 3.98E-02 | Luminal |
| BOSUTINIB | SRC (basal) | 1.08E-02 | Basal |
| DOCETAXEL | TUBB1 (basal) | 1.27E-02 | Basal |
| BMS.536924 | IGF1R (basal) | 4.95E-02 | Basal |
| VX-680 | JAK2 (basal) | 4.95E-02 | Basal |
| ERLOTINIB | EGFR (basal) | 2.33E-02 | Basal |
| RDEA119 | MAP2K1/MEK12 (luminal) | 2.04E-02 | Basal |
| TCS 2312 DIHYDROCHLORIDE | CHEK1 (luminal) | 1.46E-01 | Not significant |

**Figure legends**

**Figure 1. Analysis of Integrated Networks for Breast Cancer Cell Survival and Proliferation.**
The data of genome-sequencing, genome-wide RNAi screening, copy number variations and gene expression profiles of individual lines were used for constructing an integrated network for each individual cell line. Cell line-specific networks across each of the breast cancer subtypes were used for constructing subtype-specific networks for cancer cell survival and proliferation. Comparative and differential analysis of the subtype-specific networks allowed to predicting subtype-specific treatments and also significantly classifying breast tumor samples. See also Figure S1 and Table S1.

**Figure 2. Hierarchical Clustering of the 16 Breast Cancer Cell Lines**
Hierarchical clustering of the cell lines using cell line-specific network hubs: (A) driving-regulator hubs, (B) essential gene hubs, (C) The hubs of essential genes and driving-regulators combined. Red and beige in the heatmaps indicate whether the hub genes are present or absent, respectively, in a cell line. See also Table S3.

**Figure 3. Subtype-Specific Survival Signaling Networks for Basal A (A), Basal B (B) and Luminal (C) Subtypes.** Nodes represent genes while links represent regulation (directed links) or interaction (neutral links) between genes. A node is represented by a pie chart that shows each gene's distribution as essential gene (red), a driving-regulator (blue) or a proliferation influencing gene (cream) in its subtype. The background color behind the clusters represents a cluster's function in relation to one of the cancer hallmarks: apoptosis (pink), cell proliferation (green) and cell cycle (blue). Cytoscape (Saito et al., 2012) was used to present and visualize the networks. See also Figure S2 and Table S4.

**Figure 4. Clustering of 16 Breast Cancer Cell Lines and 402 Breast Tumor Samples Using the Hubs from Subtype-Specific Networks**
(A) Hierarchical clustering of the 16 cell lines using the differential hubs from the subtype-specific networks of luminal and basal subtypes. In the heatmap, for a given cell line, if a hub gene is an essential gene, a diving-regulator or a proliferation influencing gene, it appears in red, otherwise in beige. On the side bar, grey and yellow represent luminal and basal cell lines, respectively.
(B) The same differential hubs from A were used to classify 402 breast tumor samples. In the heatmap, red represents mutated genes or amplified genes that are among the top 50% of the expressed genes for tumor samples.

Fig 1

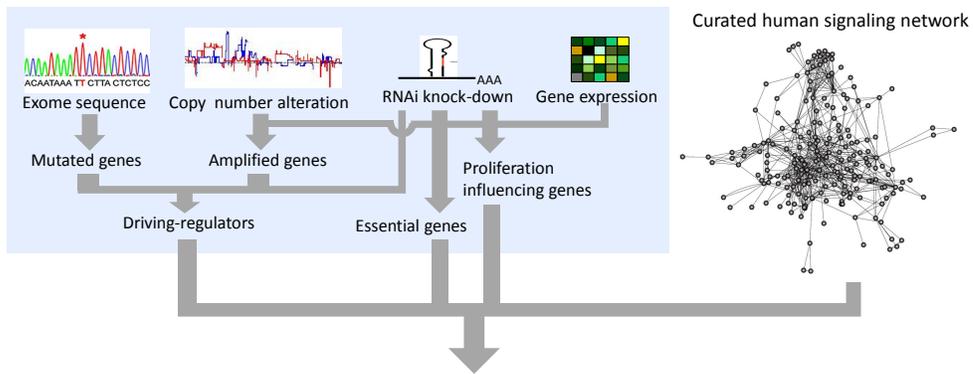

**Construct Cell Line-Specific Network**

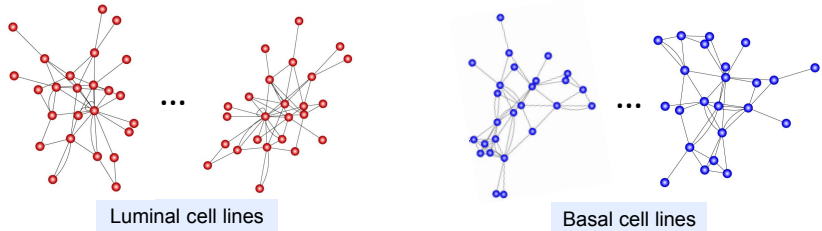

**Construct Subtype-Specific Network**

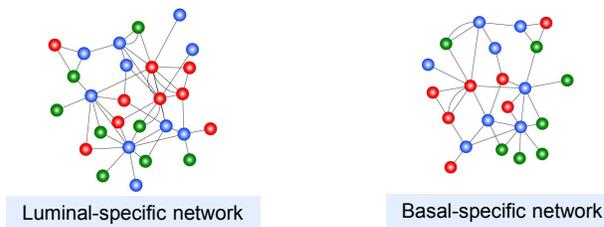

**Comparative Network Analysis, Prediction and Validation**

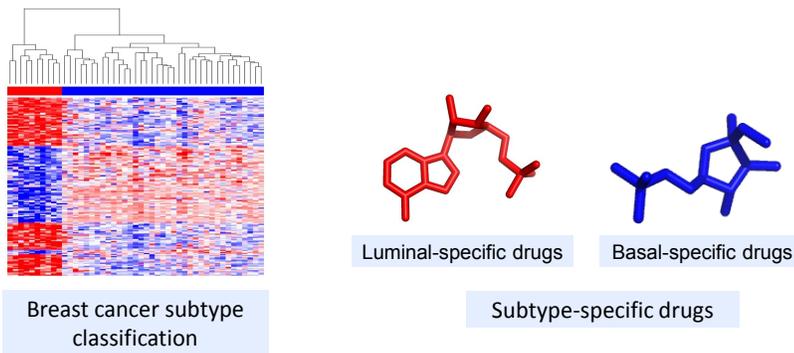

Fig 2

A B C

Fig 3

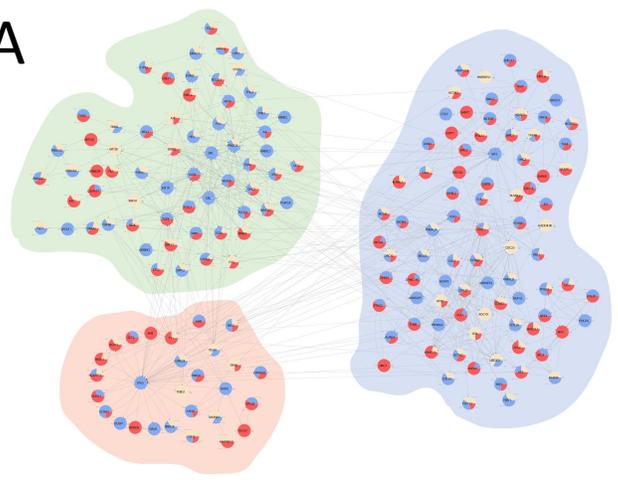

A

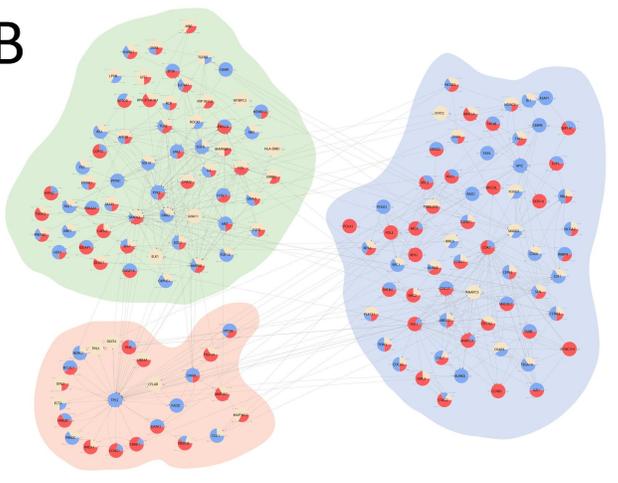

B

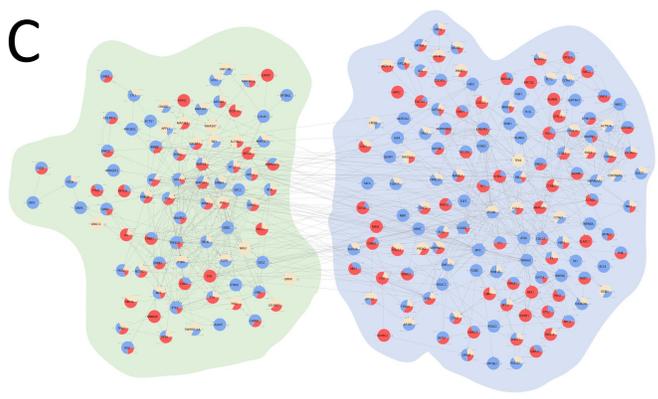

C

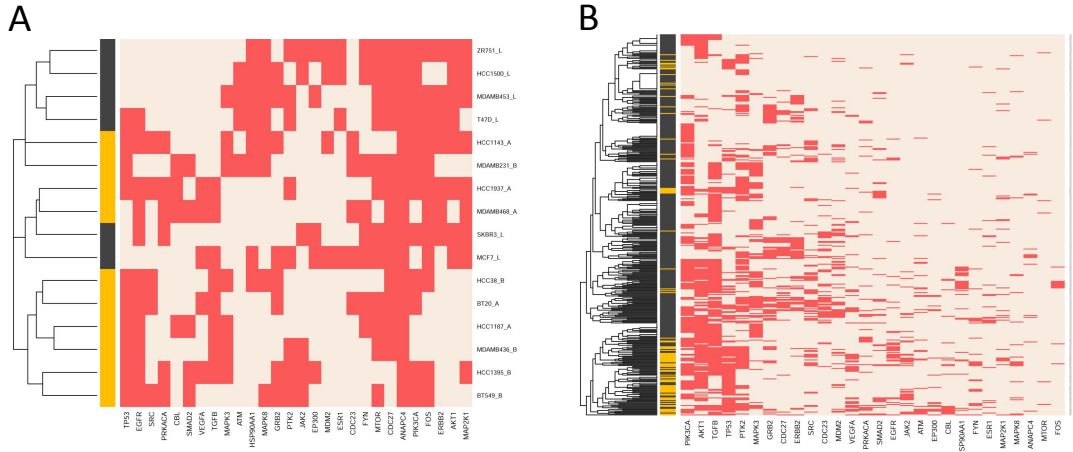